\documentclass{sig-alternate-05-2015}

\begin{document}

\setcopyright{acmcopyright}





%
\conferenceinfo{SEXI}{2016, San Francisco, USA}

\title{Efficient filtering of adult content using textual information}
%
%
%
%
%

\numberofauthors{3} 
%
\author{
%
%
\alignauthor
Thomas Largillier\\
       \affaddr{Normandy University}\\
       \affaddr{Caen, France}
\alignauthor
Guillaume Peyronnet\\
       \affaddr{Nalrem Medias}\\
       \affaddr{Paris, France}
\alignauthor Sylvain Peyronnet\\
       \affaddr{Qwant \& ix-labs}\\
       \affaddr{Rouen, France}
}

\maketitle
\begin{abstract}
Nowadays adult content represents a non negligible proportion of the Web
content. It is of the utmost importance to protect children from this
content. Search engines, as an entry point for Web navigation are
ideally placed to deal with this issue. 
  
In this paper, we propose a method that builds a safe index
\textit{i.e.} adult-content free for search engines. This method is
based on a filter that uses only textual information from the
web page and the associated URL.
\end{abstract}



%
%

%
%


\keywords{Adult content filtering}

\section{Introduction}
Protecting youth from  adult content on the web is an important
issue. A study by Sabina {\em et al.} \cite{sabina} shows  
that 93\% of boys and 62\% of girls are exposed to online pornography
during their adolescence. The mean age at first exposure to adult
content is around 14. 

It is important to filter adult content for many reasons.
We see in \cite{sabina} that not all exposure to
pornography are on purpose. More precisely, almost 7\% of boys and
42\% of girls answering to the study stated that they never looked for
pornography on purpose. We also see in this study that the exposure to
deviant sexual activity and child pornography (which the viewing
is illegal) is far from being negligible. There is thus a real issue
about filtering pornography, and more generally adult content. It is
also worth noting that it is legally forbidden in most countries 
to allows or facilitate the access to pornography to minors.

To ensure a safer Web experience search engines decided to offer a
\textit{safe search} option which objective is to remove adult content
from search engines results pages.

In this paper, we propose a methodology to construct an adult content
free index for a search engine. This index leads to a \textit{safe
search engine} rather than an option that can be deactivated.

Our methodology consists in an algorithmic pipeline which main
element is a decision forest generated by a supervised machine
learning algorithm. 

We chose to construct a Web index without any adult websites rather
that flagging in a standard Web index websites containing unsafe
content for a specific reason.  Indeed, when a user enters a query
into a search engine, the pages that are output as relevant for this
query are those that have high popularity (in term of pagerank for
instance) and that have a content relevant to the query.  By removing
most of the adult websites from the index, we nullify the popularity
of the remaining ones (websites tend to link themselves mostly if they
are in the same topical cluster). Moreover, our filter is based on the
textual content of the websites, meaning that the adult websites that
are missed by the filter does not have an enough adult content to rank
on adult queries.  This approach is thus interesting for several
reasons: it is efficient, it is fast and the few false negatives are
demoted in the search engines results pages, so most of the time
nobody will see them.

The structure of the paper is as follows, section~\ref{related}
presents the related work. In section~\ref{fast} we give the general
architecture of our filter. In section~\ref{decision} we describe our
methodology, the attributes we analyse on Web pages and the
experimental results we obtained.

\section{Related work}\label{related}
Identifying adult content on the Web is an active topic since the
democratization of the Web. It is a problem of the utmost importance
since children have an ever easier access to online resources that was 
extensively addressed in the literature.

Most techniques focus on detection of adult content detection in media
like images. Chan \textit{et al} introduce in \cite{chan1999building}
the idea of using skin related features in images to detect
pornography. However this first step prove efficient to detect images
containing a lot of ``skin'' pixels but was not precise enough to
efficiently detect pornographic content.

Rowley \textit{et al} present in \cite{rowley2006large} the filtering
system used in Google at the time. Their objective is to provide a
very fast method since the volume of data they have to classify is
tremendous. They train a SVM on 27 features based on skin and form
detection. Their results are not spectacular but they are working on a
real life dataset of over 1 billion images that is really difficult.

Hammami \textit{et al} develop
Webguard~\cite{hammami2004adult,hammami2006webguard}, a tool for
filtering adult content on the Web. Webguard uses textual, structural
and visual information on a Web page before taking a decision. It was
the most complete tool at the time and it outperformed all its
competitors. The authors show in~\cite{hammami2006webguard} that
textual and structural information already do a fantastic job at
detecting adult content. Visual information only being used to detect
false negative and improving the efficiency of the method by a
margin. In their papers the authors give no indication on the
performances of their tool.

Jansohn \textit{et al} in~\cite{jansohn2009detecting} focus on
detecting adult content in videos. Their approach consists on working
on images extracted from the video as well as motion features
extracted from the video. Using motion histograms together with a bag
of visual approach yields very good results for classifying videos.

The following two papers have the same objective as ours, protecting
children from pornography. They both focus on adult content accessed
using mobile devices.

Amato \textit{et al} introduces a parental control tool that tests
images received on a mobile device before granting access to it.
Their method intercepts images notification and transfer the image to
a remote server that can classify the image before returning the
result. If the image is inoffensive the notification is put back in
the mobile device queue, if the image is not suited for a child it is
simply deleted and the user wont even know he has received offensive
content. Their process rely on a existing image classification system
and runs in seconds per picture which is totally untractable for
running during indexation.

Park and Kim proposes in~\cite{park2010harmful} an authentication
system to access restricted content in Mobile RFID service
environment. Their system proposes a better anonymity for users as
well as a better protection for minors. The proposed system only takes
into account the access to restricted and requires a already
classified collection of content.

\section{Fast filtering of adult content}\label{fast}
\begin{figure}[h!]
 \centering
 \includegraphics[width=\linewidth]{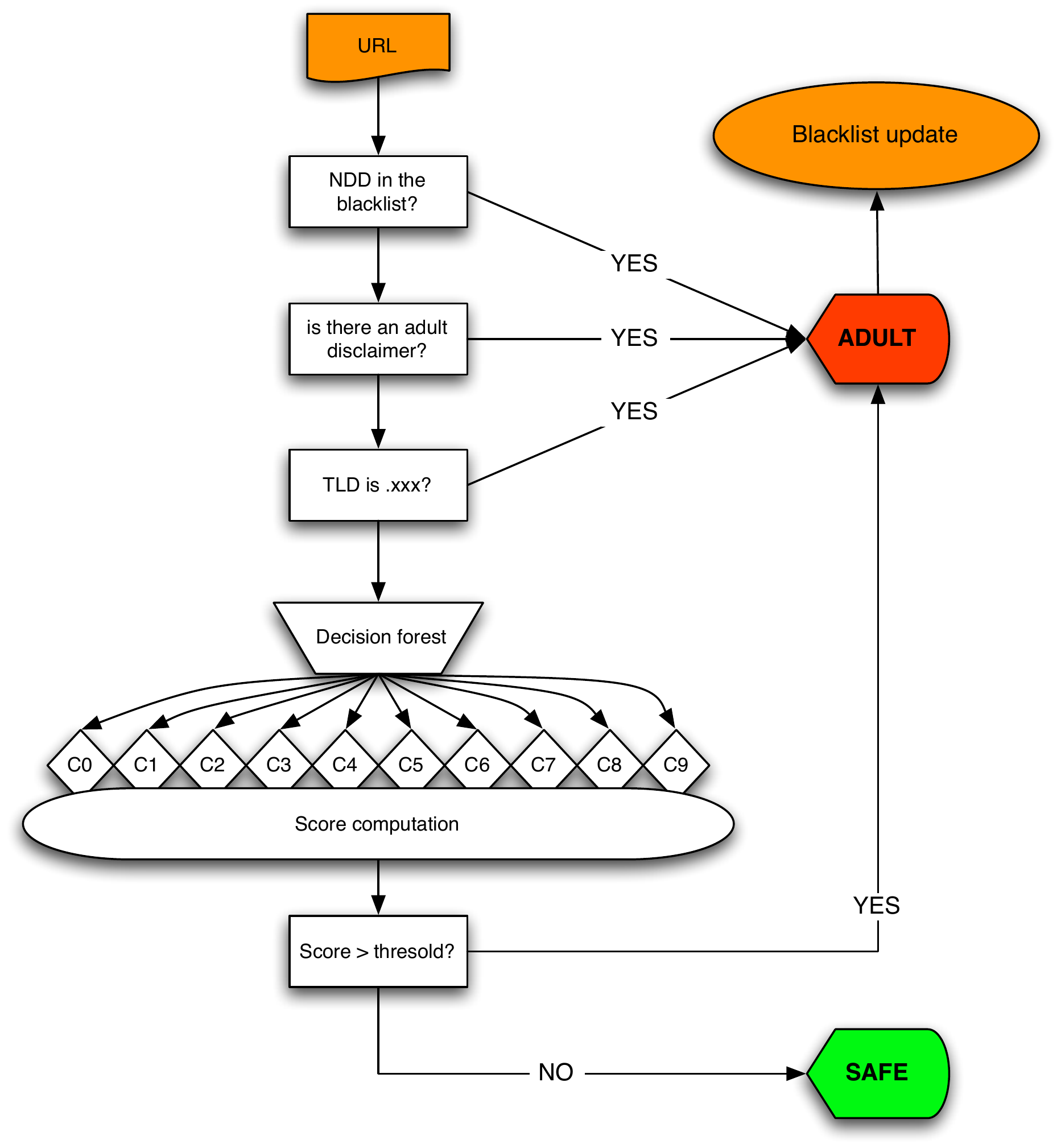}
   \caption{Principle of the filter\label{scheme}}
\end{figure}

Fig.~\ref{scheme} depicts the principle of our fast filter. 
The goal of the filter is to construct a search engine index free of
adult content. It is impossible to guarantee that there won't be any
false negatives, meaning that there will still be some Web pages
containing adult content in the final index. However, if the filter
remove, for instance, 98\% of adult content, we claim that a search
engine using this index will almost certainly never show adult content
to its users. Indeed, with only very few adult websites in the index,
the pagerank of those websites will be low, meaning that to be in top
position, a website will need a very relevant content. As we see in
section~\ref{exp}, the filter fails mainly on websites without
content, so it is unlikely that the index contains adult website with
relevant content.

We now describe the global architecture of the filter.\\

\noindent{\bf Blacklist.} To have the most efficient filter, we use a
{\em blacklist} mechanism.The first step is thus to chack whether or not the
Web page under analysis is in the blacklist.

\noindent{\bf Adult content disclaimer.} For legal or moral reasons,
most adult websites declare themselves as such using a
disclaimer. Those sites are thus easy to filter.

\noindent{\bf TLD is ``.xxx''.} Websites whose TLD is ``.xxx'' contain
adult content.

\noindent{\bf Decision forest.} The main part of the filter is a set
of decision trees obtained using a statistical classifier. In this
paper, we chose to use the C5.0 algorithm, an improved version of the
well-known C4.5 (see \cite{quinlan2014c4} and \cite{c5} for more information).
The score of a candidate page (for indexing) is the number of decision
trees concluding that the page contains adult content. If the score is
higher than 50\% then the page is considered as unsafe.

\noindent{\bf Blacklist update.} The blacklist is automatically
updated as follows: if 3 pages from the same website are considered
unsafe, then the domain name of the website is inserted into the blacklist.

\section{Decision forest}\label{decision}
In this section we first present the methodology we use to train a
decision forest to classify Web pages as safe or unsafe
(e.g. containing adult content). We then describe the attributes used
by the classifiers. Finally, we present experimental
evidence that our approach gives satisfying results.

\subsection{Methodology}
We used the C5.0 algorithm of Quinlan~\cite{c5} in order to obtain
a decision forest that allows for the classification of the content of
websites. This means that we obtain several independent decision trees
(10 in our case) that will be used concurrently in order to obtain
more accurate classification results.

To construct several decision trees, we used the {\em boosting} option
provided by the C5.0. Moreover, our goal is to have a filter with a
low percentage of false negatives (adulte websites considered as
safe), while false positives are less harmful (safe websites considered
as containing adult content). Therefore, we penalize the false negatives by
making then 20 times more costly than false positives in the C5.0
iterations.

The dataset used to train the decision forest is composed of 226 Web
pages, 120 of them being from adult websites. These websites were
chosen manually and are representative of adult websites
(youporn-likes sites, discussion forum about pornography/sexuality, 
swinger listings, erotic fiction and sex stories).  Note that creating
such a dataset is a matter of trial and error, and this step cannot
really be automated.

Once the dataset is created, we gave it as input to the C5.0
implementation of Ross Quinlan, together with features extracted from
an analysis of several attributes over pages from the dataset.

\subsection{Description of the attributes}\label{attri}
We chose to use only attributes that can be analysed quickly. This
means that they are internal to pages and belongs to either textual
content or quantitative attributes (e.g. number of images).
We now describe each attribute and the associated computations.\\

\noindent{\bf in-url.} We check the URL for the presence of given
terms. We decided to use a list of 27 terms, from generic
word (``porn'') to brands (``cam4'', ``tube8'', etc.).\\

This attribute is used in two ways. We compute the number of terms
from the list that are in the domain name, but also the number of
terms that are in the URL. Moreover, we are looking for substring,
meaning that, for instance, ``sex'' is considered as being in
sexhungrymoms.com.

The following attributes are also defined by a list of terms.
Each being subject to 3 different computations. We first compute the
number of occurrences of terms from the file X in the page (denoted
nb\_X in the following). Then we
compute the proportion of words from the file in the page (denoted
ratio\_X). Last, we
make the reciprocal computation, that is the proportion of words from
the page that are also in the file X (denoted
prop\_X). \\

\noindent{\bf brand-names.} A set of 34 brands related to adult
content (content producers, major websites, etc.).\\

\noindent{\bf categories-en.} 222 english terms that are usual categories of
pornographic websites.\\

\noindent{\bf categories-fr.} 593 terms used by french pornographic websites 
as categories (terms can be in french or english).\\

\noindent{\bf categories-gen.} 79 french categories.\\

\noindent{\bf en-words.} 100 english terms for ultra-sensitive topics
(child pornography, rape, etc.).\\

\noindent{\bf french-words.} A list of 163 french words representative
of adult content (the goal of this list is to filter erotic litterature).\\

\noindent{\bf pornstars.} Names of 8825 adult entertainment industry
actors (male and female).\\

\noindent{\bf queries.} 716 typical adult queries (e.g. ``porn
gratis'', ``porn gallery'', etc.).\\

\noindent{\bf small-set.} 11 terms that are common on adult websites
(e.g. ``sex'', ``xxx'', ``porn'', etc.).\\

\noindent{\bf tags-en.} 2000 most frequent tags (in english) for
pornographic videos.\\

\noindent{\bf tags-fr.} Similar to the previous list, 69 tags in french.\\

The last attribute is quantitative, and concerns images.\\

\noindent{\bf \# images.} Number of images in the page.\\

One could remark that we are not using the number of videos or ads as
an attribute. Our experience is that it does not give better results to
use these attributes, and they are more difficult to compute than all
the attributes above.

\subsection{Obtained Decision trees}\label{classifiers}
\begin{figure}[h]
 \centering
 \includegraphics[width=.8\linewidth]{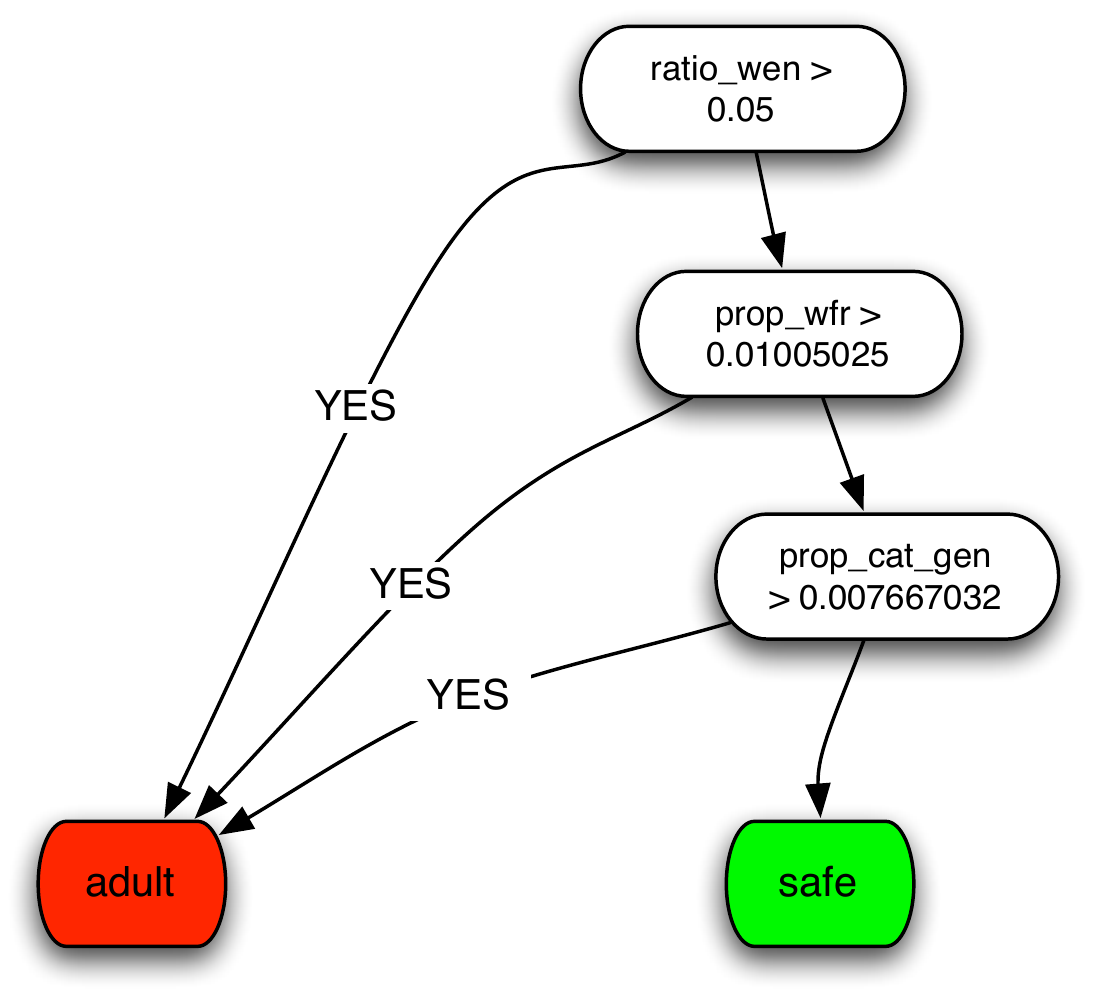}
   \caption{Decision tree \#2}\label{3DT}
\end{figure}

We obtained 10 decision trees, whose errors on the training dataset
are depicted in the Tab.~\ref{tab:DT}.
\begin{table}
  \begin{center}
    \begin{tabular}{|c|c|c||c|c|c|}
      \hline
      tree id & size & error & tree id & size & error \\
      \hline
      0&	     7&   13.7\% &  5&	     6 &   7.1\%\\
      1&	     3 &  23.0\% & 6&	     8 &  12.4\% \\
      2&	     4  &  9.7\%  & 7&	    10 &    3.5\% \\
      3&	     6   & 6.2\%  &   8&	    11 &    3.5\% \\
      4&	    10 &   8.0\% & 9&	    10 &    3.5\%\\
      \hline
      \hline
      \multicolumn{3}{|c|}{global error} & \multicolumn{3}{|c|}{ {\bf 0\%}}\\
      \hline
    \end{tabular}
  \end{center}
  \caption{Decision trees sizes and errors\label{tab:DT}}
\end{table}
We present in Fig.~\ref{3DT} a graphical view of the decision tree \#2
(the one with a $9.7$\% error).
This tree has a small complexity since only 3 attributes are used. When
a Web page is analyzed, it first computes the proportion of terms from
{\bf en-words} that are in the page (noted ratio\_wen here). If
ratio\_wen > 5\% then the page is considered as unsafe, otherwise the
proportion of terms in the page that are also in {\bf french-words} is
computed, and compared to a threshold. The page is filtered depending
on the comparison. Last, the attribute {\bf categories-gen} is used.

All decision trees behave similarly, each of them being important to
the final decision. Each decision tree has its own (sometimes high)
error percentage. It is the aggregation of the decision of all trees
through a majority vote that allows to a small global error.

We use a majority vote to aggregate decision from all
trees. However, it is possible to tune the filtering power of our
method by lowering the threshold (for instance, we can consider that a
Web page is unsafe if at least 3 decision trees are in agreement on
this outcome). Lowering the threshold will increase the number of false
positive, which is less harmful when dealing with unsafe (adult)
content.

In section~\ref{attri}, we saw that we deal with 36 different
attributes. Amongst them, 23 are used by the decision forest. But not
all 23 are used for analyzing each web page. 6 attributes were always
computed when classifying Web pages from the training dataset:
ratio\_cat\_gen, prop\_cat\_gen,\\ \mbox{prop\_wfr}, prop\_small,
prop\_queries. All these attributes depend on the presence of very
specific terms in the textual content of the page. Both ratio and
proportion are important, which is natural since it is unlikely that
an adult website contains only one category of adult content, or only
a few terms relevant to adult topics.  Other attributes are used with
frequencies presented in the following table.

\begin{table}
  \begin{center}
    \begin{tabular}{|c|c||c|c|}
      \hline
      frequency & attribute & frequency & attribute\\
      \hline
      73\% & IN-URL &50\% & prop\_brand\\
      73\% & ratio\_brand &  49\% & ratio\_tags\_fr\\
      70\% & ratio\_queries &   46\% & IN-NDD\\
      68\% & ratio\_wfr &   28\% & ratio\_tags\_en\\
      65\% & nbr\_img &  28\% & nb\_star\\
      63\% & ratio\_star &  24\% & nbr\_brand\\
      58\% & prop\_cat\_en &  24\% & nb\_wen\\
      53\% & nb\_wfr & 19\% & nb\_tags\_fr\\
      51\% & prop\_star &  &\\
      \hline
    \end{tabular}
  \end{center}
  \caption{Attributes usage frequency\label{tab:auf}}
\end{table}
We learn from Tab.~\ref{tab:auf} that an attribute such as the domain
name is more useful at the blacklist level than at the decision forest
level. Indeed, major players from the adult entertainment industry
have domain name that are brands and as such that does not contain
typical adult terms.
However, attributes about the URL are important, mainly because many
adult websites are using tags and categories in the URL in order to
improve their search engines optimization (SEO). This is for instance
the case of \textit{youporn.com}.

\subsection{Experiments}\label{exp}
We saw in Sec.~\ref{classifiers} that we obtain
very promising results on the training dataset. However, it does not
means that the filter will be efficient on new data.

So we tested our filter on a new sample of Web pages. This
dataset is composed of 1153 pages, 839 contain adult content, and 314
are totally safe.
The results are summarized in Tab.~\ref{tab:res}.
\begin{table}
\begin{center}  
\begin{tabular}{|c|c||c|}
  \hline                
 (a) & (b) & <- classified as\\
  \hline
 821  &  18 &   (a): class adult\\
	    14 &  300  &  (b): class safe\\
    \hline
\end{tabular}
\end{center}
        \caption{Classification results\label{tab:res}}
\end{table}
These are satisfying results. The miss rate is (all the following
numbers are rounded values) $2.15 \%$, the accuracy is $97.22 \%$, the
recall is $97.85 \%$ and the precision is $98.32 \%$.

Note that we have tested here only the decision forest of the
filter. We did not used the blacklist update mechanism, nor the TLD
and disclaimer detection. If we add these additional mechanisms, the
accuracy is higher.

It is interesting to understand why some adult websites are considered
as safe by our filter (false negatives). We made 3 additional tests,
with specific types of adult websites.
What we learned from these tests is that some adult video streaming
website contains images and videos without explicit textual
content. This is for instance the case of the website {\em beeg}, a
youporn-like website. Since our filter is looking only at textual
content, it cannot detect such website.
A more ambiguous example is the one of discussion forums. While some of
these forums are very explicit (this is for instance the case of
``swingers forums''), others are forums of classical websites with a
``sexuality'' section. On those very specific websites, our filter is
performing poorly. This is however a minority of the pages containing
adult content, and the content of these forums is mildly explicit (no
images, no videos).

\section{conclusion}

In this paper, we presented a method that builds a safe index for
search engines. Our experiments show that the method is efficient.
Using only textual content proves sufficient in almost all
cases with an accuracy of $97.22\%$.

\noindent\textbf{Aknowledgements}. The authors would like to thank
Qwant for funding parts of this research.

%
\bibliographystyle{abbrv}
\bibliography{SEXI}  
%
%

\end{document}